# A Brief Perspective on Piezotronic and Thermoelectric Coupling: Flexible Platforms for Synergistic Energy Scavenging and Peltier-Caloric Effects

David L. Carroll
The Nano/Quantum Technology Laboratory (NanoteQ)
Department of Physics
Wake Forest University
Winston-Salem NC 27109 USA

ChaoChao Dun
Lawrence Berkeley National Laboratory
Berkeley, CA 94720

**Abstract**

Advances in the development of flexible piezoelectric and thermoelectric materials have provided an important avenue for the exploration of energy scavenging through the thermodynamic-coupling of orthogonal energy-scavenging modalities. This has led to a body of work creating hybrid thermo/piezo-electric generator devices (T/PEGs) in which the two effects become thermodynamically entangled. Based on hetero- thin film architectures, such devices can exhibit a surprising power generation characteristic which is non-additive between the two energy-scavenging effects. For example, when the thermoelectric and piezoelectric effects are strongly coupled by their proximal fields, the efficacy of energy scavenging can be made to exceed that of the two effects independently. In this review of such effects, a basic coupled heat engine model is shown to provide insight into the origins of synergistic power generation. These models however, suggest the emergence of other combined thermodynamic properties such as the kinetic Peltier-Caloric Effect (PCE) traced to Onsager reciprocity. The recent observation of this effect in multicomponent systems is confirmation of the limited perspective of thermodynamic separability.

**Outline**

1. Energy scavenging and power generation
2. The Thermodynamics of Combined Scavenging
   2.1 Combining scavenging modalities as a thermodynamics problem
   2.2 Entangling Modalties
3. Combining Thermoelectrics with Piezoelectrics: the T/PEG
   3.1 Combining Modalities Through Proximal Fields
   3.2 Measurement Techniques for Planar T/PEGs
4. Materials
   3.1 SWNT Matrix Nanocomposites
   3.2 Chalcogenide Matrix Nanocomposites
5. Scavenging Power to Cool: Combining Peltier with Electrocalorics
6. Conclusions

## 1.0 Energy scavenging and power generation

Ambient waste heat is notoriously difficult to scavenge and convert into useful power. This fact is well studied and limits on scavenging are well quantified by various thermodynamics principles and models. This means that one significant focus of research in the field of ambient power scavengers has turned to platforms that more effectively allow access to heat reservoirs, not just on the improvement of local efficiencies. Thus thin, low cost, light weight, and large-scale coverings or coatings can be just as effective in generating power as a small but highly efficient strategically-placed thermoelectric or piezoelectric device.

## 2.0 The Thermodynamics of Combined Scavenging

The human body produces ~ 100 W of waste heat in unloaded metabolic operation. But this waste heat is clearly not all. Motion such as walking, breathing, shivering, along with the useful work done, also produces some waste from the body's perspective. Indeed, many machines, engines, or generators are thermal sources and generate some useless vibration. Think of the heat and vibration on the tailpipe of a car while running. So, the volume of space around the tailpipe contains wasted heat energy from the engine combustion and the vibrational energy of that engine, road vibrations, etc. Expanding access to the waste energy reservoir in these examples simply means collecting both the thermal and kinetic waste energy. In terms of solid-state physics that means combining piezoelectrics with thermoelectrics for this example.

### 2.1 Combining scavenging modalities as a thermodynamics problem

Of course, an obvious claim is: if there is waste heat, vibration, radiation, or friction, generated by some system, then collecting it all is good. In simplest terms consider the thermodynamic behavior of a multi-mode scavenging system as the summed behaviors of individual energy scavengers: heat, light, vibrations and so on. So, the total energy scavenged in such a system is simply:

$$E_{total} = \int \sum_i A_i \eta_i P_i \, dt$$

Where $A_i$ is the effective interface with the reservoir (ie. The access to the waste heat), $h_i$ is the efficiency of transfer to useable energy, $P_i$ is the total local power within $A_i$. Each of these entities can be time dependent. The index *i* runs over the different modes of scavenging the system can accomplish. This result is not very exciting, it simply says that if a thermoelectric and a piezoelectric are strapped to the same bit of machinery, the power from the combination is simply the sum of the two. This additive approach drives up the costs of scavenging with each device added and the whole purpose is to retrieve waste power at a low-cost.

### 2.2 Entangling Modalities

It is easy to imagine a situation where the scavenging of energy in one form in some environment might influence the collection of energy in another form. For example, consider a Si photovoltaic with a thermoelectric system on the back. Such a stacked dual device can collect both > 1 eV photons, and the heat generated by the IR part of the spectrum. However, by removing the IR derived heat from the module, the PV is cooled by some small amount. Thus, the actual efficiency of the PV process in enhanced slightly. That is, more power is generated than if the two devices were deployed separately. A second example might be an organic bulk heterojunction photovoltaic. [1] These solar cells are fabricated with a nanophase dispersed throughout a polymer photon absorber as in **Figure 1**. Normally this is some kind of p-type thiophene mixed with a $C_{60}$ derivative to act as an electron acceptor. But what if the materials are chosen so that the nanophase is also an p-type thermoelectric material? So, we put the thermoelectric *inside* the photovoltaic. Then, under solar flux, a ΔT from top to bottom of the film will result. This yields internal potentials which would be small, but collectively they aid electron drift, improving the overall performance beyond what might be gained from the photovoltaic power alone.

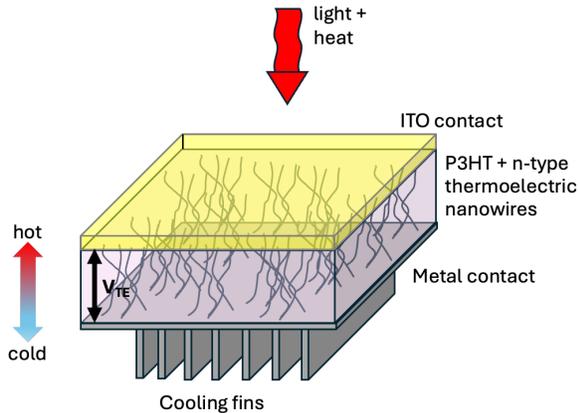

**Figure 1:** The example of combining modalities in PV. We can collect the heat of IR in a bulk heterojunction photovoltaic by selecting the nanophase as an p-type thermoelectric material. The DT is established by absorbing IR and cooling the back of the device. This heat creates an internal thermovoltage ($V_{TE}$) to aid electron drift through the matrix. This simple thought experiment suggests that the overall principle is widely applicable.

As a third example, consider a thermoelectric material that is layered with a piezoelectric material and the field from the distressed piezo penetrates the thermoelectric thereby coupling the two. This is the topic of the current review. The penetration of the piezotronic field into the thermoelectric can be aided by using matrix composites in which the host material is a dielectric. Under specific conditions, this field can be used to do work within the thermoelectric increasing its scavenging efficiency in analogy with the other two examples.

What general approach should be used to address "coupled" systems thermodynamically? Does the coupling give more or less power than the simple combination? More precisely: given two proximal heat engines that attach to two separate thermal baths, is maximal work extracted by using the heat engines separately or by allowing heat engine 2 to do work on heat engine 1 thereby increasing heat engine 1's efficiency to do work?

If we examine the power term in Equation 1:

$$P_{total} = \sum_i A_i \eta_i P_i$$

Consider the case wherein the efficiency of one modality of scavenging is parametric in the others:

$$\eta_j(\eta_1, \eta_2, \eta_3, \dots \eta_{j-1}, \eta_{j+1}, \dots \eta_n)$$

As a simple first order approximation, we apply a linear model from a Taylor expansion. In a multi-variable system, we get:

$$\eta_j = \alpha_j \eta_j + \sum_k \sum_{i \neq j} \alpha_i \eta_i^k$$

Where 0 < k £ n and the a's are the appropriate derivatives. For two variables:

$$\eta_1|_{coupled} = \alpha_1 \eta_1|_{P_2=0} + \alpha_2 \eta_2|_{P_1=0} = \alpha_1 \eta_{1o} + \alpha_2 \eta_{2o}$$

So,

$$P_{total} = A_1(\alpha_1 \eta_{1o} + \alpha_2 \eta_{2o})P_1 + A_2 \eta_2 P_2 \approx A_1(\alpha_1 \eta_{1o} + \alpha_2 \eta_{2o})P_1 + A_2 \eta_{2o} P_2$$

Here we have assumed that while $\eta_{2o}$ effects $\eta_{1o}$ it doesn't go the other way. We further assume that this means $\alpha_1 \approx 1$. So, only a little of $\eta_{2o}$ is mixed into $\eta_{1o}$ to get $\eta_{1\ coupled}$. This is because we are considering only the cases where the efficiency of process 2 increases the efficiency of process 1 (as in our solar cell example). Of course, it can go either way, but this recognizes that even for the simplest case of a 2 variable, linear model, when $\alpha_1 \to 1$ and $\alpha_2\ small$ the power generated will increase by:

$$P_{total} \approx A_1 \eta_{1o} P_1 + \alpha_2 A_1 \eta_{2o} P_1 + A_2 \eta_{2o} P_2$$

and the additional work done: $\alpha_2 A_1 \eta_{2o} P_1$ depends on the coupling constant $\alpha_2$ and the power supplied by process 1 converted at the efficiency of process 2.

A more detailed treatment of this question would be handled using Onsager's formalism. [2] Indeed, when the off-diagonal terms in Onsager's matrix express cross terms between heat engines, one might more accurately refer to this problem of coupled heat engines as "Onsager entanglement." However, our simple analysis above is illustrative and gives the simple limits that can be expected for enhanced scavenging and work produced from such systems.

**3.0  Combining Thermoelectrics with Piezoelectrics: the T/PEG**

This simple thermodynamic picture of parametric coupling in efficiencies does not specify the mechanism of coupling nor does it require the specific type of modalities involved. It is quite general but, to put the concept to the test it is necessary to choose a system. In this case we will review advances in the construction of dual scavenger assemblies that access waste power through both thermoelectric effects and piezoelectric effects. In fact, there has been a developing focus among researchers in the field to combine kinetic energy scavenging and thermal energy scavenging on the same body with a single monolithic platform. However, large-scale, fabric-like conformal coverings, clothing, sound dampeners, and thermal insulation layers which scavenge from these two sources has proven a challenge in terms of power/cost ratios and in terms of lifetimes. This is because thermoelectric materials that perform well at body temperatures typically use $Bi_2Te_3$, $Bi_2Se_3$, $Sb_2Te_3$ derivatives, which are expensive to synthesize and to process. Poled thin film, piezo-materials are also expensive and susceptible to de-poling under excessive and repeated stress. Moreover, the flexible and conformal analogues of such materials usually do not offer the performance metrics of their ceramic counterparts.

## 3.1 Combining Modalities Through Proximal Fields: the T/PEG

Organic matrix nanocomposites (OMNs) are typically made up of a nanophase of low-dimensional materials like nanowires, or single walled carbon nanotubes (SWCNTs) finely dispersed within an organic matrix – like halide containing dielectric polymers. Such polymers are known to improve dispersion and to provide some control over the nanophase morphology within the matrix. This means that above percolation, the conductivity can be controlled by loading and morphology. Since the conducting components are distributed throughout the volume, field penetration into such a volume is also quite controllable. So, OMNs offer unique opportunities in coupling thermoelectric and piezoelectric materials systems with proximal electromagnetic fields.

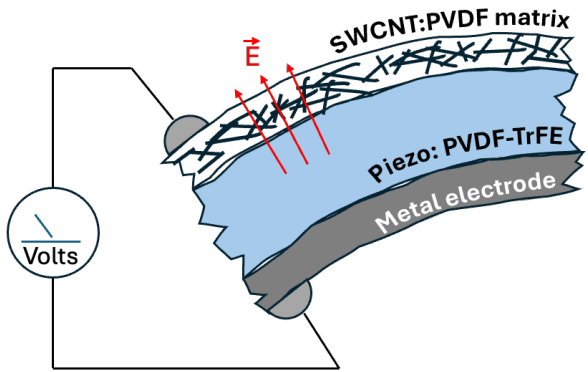

**Figure 2:** If the top, thin film electrode is made from a matrix composite of single walled carbon nanotubes (SWCNT) just above the percolation threshold, any field applied to it will penetrate somewhat into the volume of the film. When this film is deposited upon a poled piezoelectric polymer such as PVDF-TrFE, bending the stack creates a field that is then placed across the nanotubes of the upper electrode matrix.

As an example, the simplest structure for coupling the thermodynamic properties of the two materials is as shown in **Figure 2**. Here a typical piezoelectric actuator assembly is created by depositing the poled piezo-material PVDF-TrFE [3] between two electrodes. However, the top electrode has been replaced with a matrix composite of single walled carbon nanotubes in this example (SWCNTs) that are loaded to just above the percolation threshold for conduction. So, this electrode is conductive, and such nanotube composites are known to make good thermoelectrics as well. [4] However, an applied electric field penetrates this matrix contact well into the volume because of the diffuse nature of the nanoconductor fillers. So, when the stack is bent or distorted in some way, a field is generated that penetrates the matrix and modifies transport within the nanotubes. This piezotronic behavior of an applied field from a piezo-electric modifying transport in a single carbon nanotube is already well established as is the field modification of thermoelectric response. [5, 6] Extending these concepts to large-scale thermoelectric electrodes of distributed nanotubes or nanowires, we get the assembly of **Figure 2**.

In the assembly of **Figure 2** the piezovoltage and the thermoelectric voltage (or their respective currents) can be collected when a thermal gradient is applied simultaneously with distortion of the layered piezo. The piezovoltage generated will be slightly lower than if we had used a pure perfectly conducting metal electrode on the top [7] and the thermoelectric current will be larger during the bend. This synergistic multi-mode scavenging "device" has come to be called a thermo/piezo – electric generator or T/PEG (pronounced TEE-peg). It collects heat and kinetic energy simultaneously. However, because of the way the piezoelectric and thermoelectric components are arranged, the behavior of the piezo can modify the efficiency of the thermoelectric.

To implement this integrated thermo/piezo, T/PEG system in a useful way, consideration must be given to how thermoelectric generators (TEGs) and piezoelectric generators (PEGs) are implemented separately. PEGs generate voltage and are usually high impedance sources.

Commercially, a thin film of poled piezo-active material has electrodes deposited onto either side and the voltage drop during distortion is measured. However, TEGs are *current sources*; they don't make a lot of voltage. To provide voltage, n-type and p-type Seebeck materials are combined into *modules*. The *module legs* add their Seebeck voltages in series as in **Figure 3**. This will be particularly important for the T/PEG application since the relative voltages of the two components will need to be of the same order to prevent one component from drawing too much current and sinking the voltage of the other (ie. they need to be impedance matched).

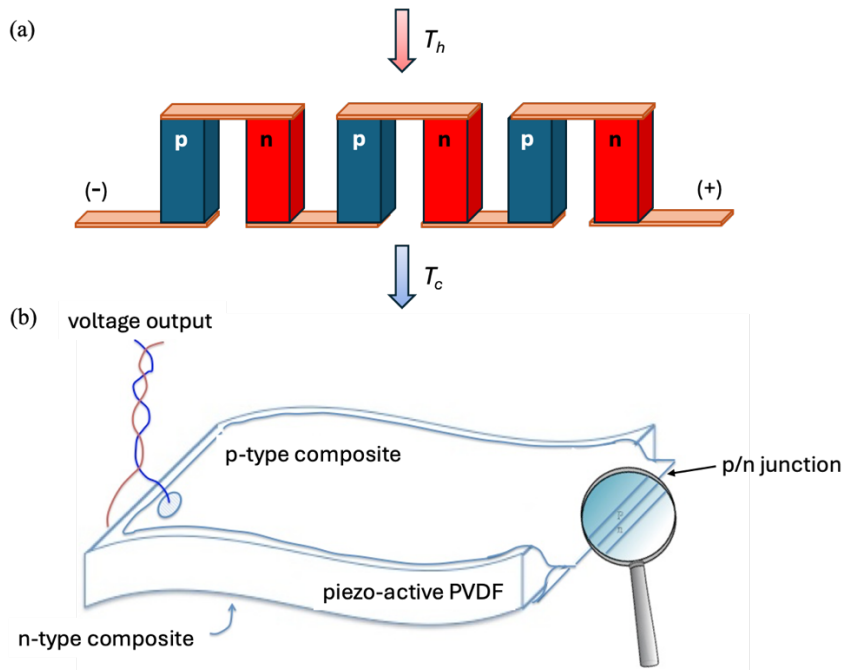

**Figure 3:** (a) The design of a T/PEG requires the use of multiple "legs" of alternating p-type and n-type matrix materials. (b) This is in analogy to a standard thermoelectric module as seen here. [8]. However, the metal interconnects of the module in (a) are replaced by p/n junctions where the p - and n - type materials connect.

**Figure 3** compares standard TE module design (a) with and arrangement of the p / n – type matrix materials of the T/PEG. Current flow is in series through the array of stacked sheets and the heat flow is in parallel to the monolith. Since this module uses flexible n-type and p-type composite thin films, a nonwoven fabric-like module results which is similar to a felt. From the perspective of *circuit topology*, these arrangements are equivalent. Of course, the DT runs laterally, which is not as convenient, but for now this doesn't matter. The usual choice of flexible piezo is PVDF (polyvinylidene difluoride) or its copolymers. [9] The legs can be any nanoscale n-type and p-type Seebeck material.

### 3.2 Measurement Techniques for Planar T/PEGs

To examine the addition of the thermoelectric and piezoelectric components under dynamical forcing (or driving) a specialized measurement platform must be used. The system should be electrically and thermally isolated of course and provision made for power collection in such a way as to allow for identification of its source. Forces and temperature gradients must be evenly applied across the relatively large test pieces that are typically used in such experiments. Note that the variations in ultrathin films used for such work make it necessary to test large samples to achieve an average output over an area. [10]

A standard testing configuration, with standard dimensions and poling etc. must be adhered to since outputs can depend sensitively to such variables. These vary across the literature and for much of the discussion in this review the T/PEG design described above is used. The assembly picture in **Figure 3** uses a p-type and a n-type thermoelectric leg deposited on opposites sides of a poled PVDF piezo layer to form one section of a simple thermoelectric module with piezo-active properties. So, the low-dimensional matrix nanocomposite material that forms the contacts to the piezo can be made of any p-type / n-type nanowires blended into some organic flexible matrix host material. A good choice is the high dielectric unpoled PVDF. This makes it easy to integrate the layered module together since the matrix material and the piezo material are compatible and adhere well.

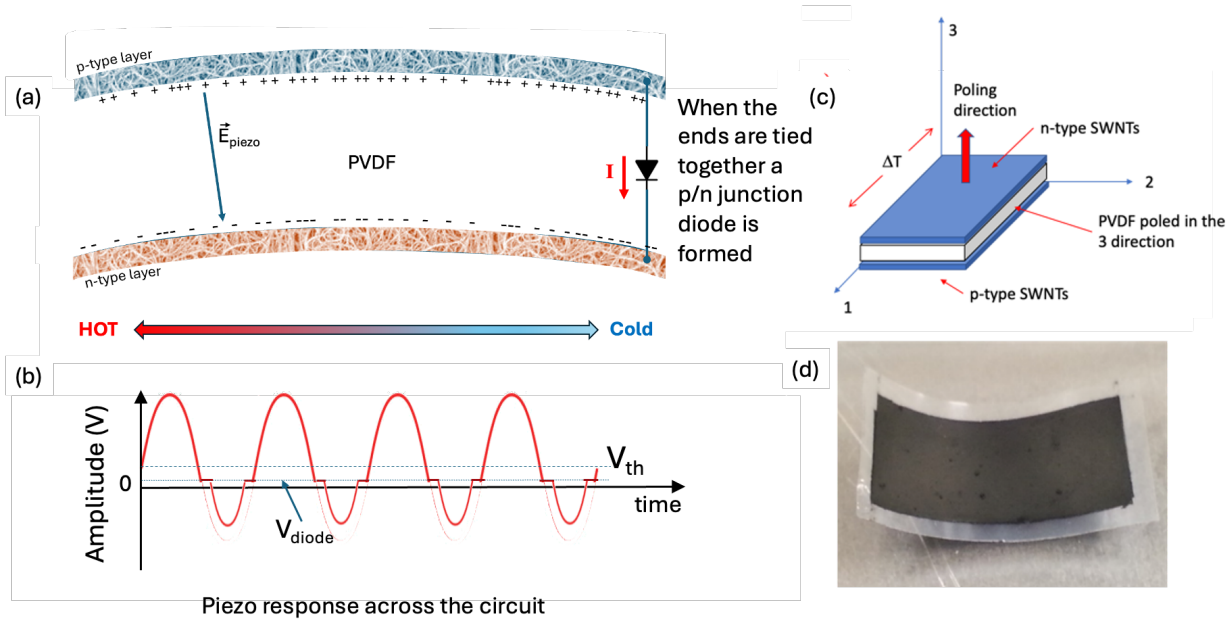

**Figure 4:** (a) Edge view of n-type and p-type matrix composite electrodes attached to a poled PVDF film of 100 microns thick. Heat flow is lateral. (b) The equivalent circuit and expected output for a three-layer stack with a DT and a sinusoidal pressure applied. The "diode" offset voltage only occurs if the p/n materials form a junction and is shown exaggerated. (c) The orientation of piezo poling relative to the plane is shown. (d) is the stack before pressing the p/n components together. This image is of SWCNT electrodes.

The thermal gradient is applied along (∥ to) the plane of the layers in **Figure 4** and the compression is perpendicular (⊥) to that plane. The $P_3$-poled layer of PVDF placed between two thermoelectrics creates a voltage that is dropped perpendicular across the plane of the thermoelectric layers. This means that the thermoelectric p-type and n-type legs allow for Seebeck conduction of charge along their length, but this conductivity is altered by the piezo-field that is applied across their thickness. In other words, a *piezotronic coupling* that will influence the thermoelectric behavior has been created.

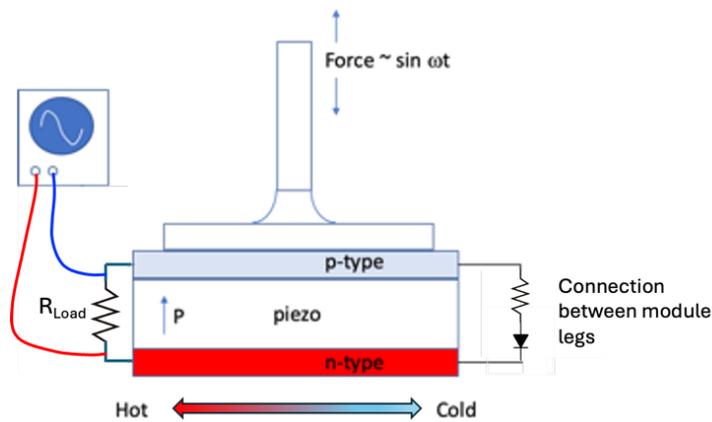

**Figure 5:** A diagrammatic drawing of the T/PEG measurements is shown. The thermoelectric and piezoelectric voltage and current (power) can be measured for an integrated system under a wide variety of conditions. Using a dynamic feedback system, thermal drift, mechanical instabilities, and other factors are compensated.

A measurement challenge of power scavenging in the activated T/PEG structure is unambiguously identifying thermal and piezo power components independently. This is needed so that second order synergistic effects might be quantified. **Figure 5** shows a setup with which isolation of pyroelectric and Peltier effects, stability of local contact heat flow, and monitoring of dimensional changes within the system can be accomplished.

This unique measurement system is isolated from the room for temperature stability and temperature ramping is accomplished using dynamic feedback. This allows pyroelectric effects to be unambiguously identified. Dimensional changes are monitored using interferometry throughout the temperature cycle because these can result in changes to internal capacitances yielding unreliable piezo measurements. Dynamical forcing in the system is continuously variable from a few Hz to kHz, over a wide variety of force values. A sinusoidal forcing function is typically used and the compression is applied evenly across the plane of the piezo stack. [11] The system is tested and calibrated using standard commercially available TEs and PEs. Schematics and pictures of the system are shown in **Figure 6** for size reference. The system itself can be thought of as a dynamical mechanical analyzer (DMA) built for thermo/piezo - measurements.

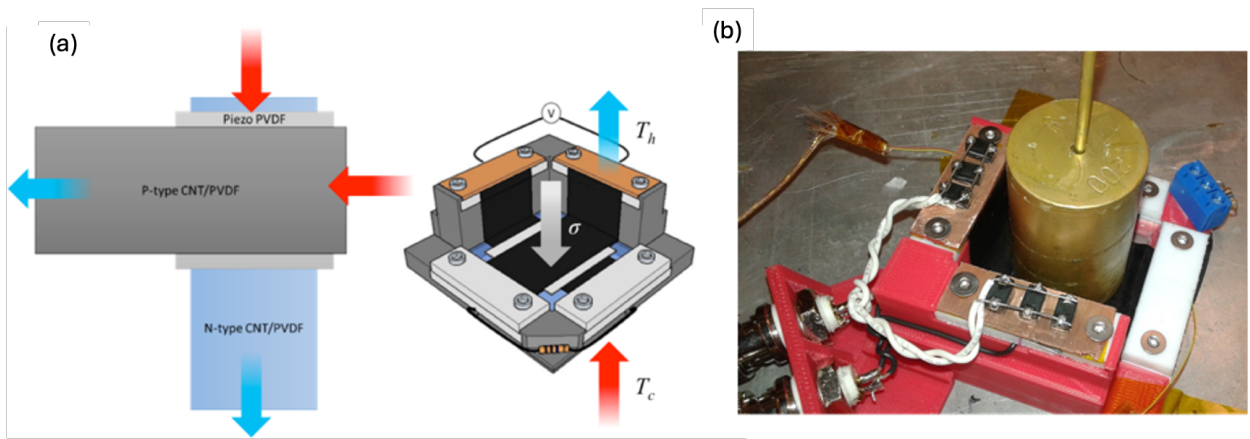

**Figure 6:** Because this is an unusual measurement, images of the apparatus are shown. This gives an idea of scale. (a) is a CAD drawing of the system. s is the plane across which the force is applied, evenly. (b) is a photo of the system. We are showing this with a three-layer T/PEG assembly. Load resistors for impedance matching, heaters, and thermocouples are all seen along the edges of the sample holder. This system is embedded within an environmental and E&M isolation box.

## 4.0 Materials

An important aspect of the performance of the T/PEG is the choice of filler material or the nanophase used in the matrix nanocomposite. Just as with nanocomposites used in thermoelectrics [12] and piezoelectrics [13] the overall morphology of the matrix blend, the electrical and thermal connectivity between nanophase components in the composites, and the mechanical freedom of the nanophase under applied stress, all play a central role in the how such matrix composites will perform. Then of course there is the central issue of the properties of the nanophase itself, its thermoelectric or piezoelectric properties, and its conductivity above the percolation threshold; that is the mechanisms of charge transport such as variable range hopping or fluctuation induced tunneling. [14]

For the purposes of T/PEGs, the basic working model above suggests the active thermoelectric matrix materials should exhibit a large Seebeck coefficient, a large thermoelectric efficiency, allow for effective field penetration of the volume and provide excellent charge transport. This is a challenge as is well known to researchers in the flexible thermoelectrics community. [15]

There are several figures of merit used to characterize a thermoelectric regardless of its material composition. Among these the most cited is the ZT = $α^2σT/κ$, where α, σ, T, and κ are the Seebeck coefficient, electrical conductivity, absolute temperature, and thermal conductivity, respectively. A second important figure of merit is the power factor PF = $sa^2$ where s is the conductivity and a is the Seebeck coefficient. The Seebeck coefficient is related to the materials properties, specifically the density of electronic states g(E) (or DOS) as:

$$\alpha = -\frac{\pi^2 k_B^2 T}{3|e|} \frac{d[\ln(g(E))]}{dE}\bigg|_{E_F}$$

## 4.1 SWNT Matrix Nanocomposites

The use of 1-to-2-micron long SWCNTs provides an excellent opportunity to test the concepts of field-doping thermoelectrics using piezoelectric fields. They can be easily p- or n- doped, they are highly conductive as a percolating matrix, and they are one-dimensional, so they have a DOS that is dominated by van Hove singular points. The thermoelectric behavior of SWCNT mats and composites near percolation relies heavily on several structural features of the matrix. Specifically, junctions between the nanotubes block phonon transfer and thus most of the heat flow. Carrier transport is dominated by 3D variable range hopping mechanisms. [16] The modification of the thermoelectric current can happen in two ways. The first is a purely piezotronic mechanism of field-doping the nanotubes. In this case the field moves the Fermi levels of the SWCNTs toward the Van Hove singular points in their band structure as seen in reference [6]. This is essentially the same as any contact dopant would do, and it raises the conductivity of the material. [17] The second modification that happens depends entirely on the bending geometry. With careful alignment of the poling and bending, a component of the electric field, internal to the matrix, can be placed in the direction of electron hopping. This increases hopping rates and conductivity if field orientation is optimal. Moreover, processing SWCNTs into matrix composites and matrix composite thin films with relatively controllable morphologies, which means control over thermoelectric and transport properties with loading, is relatively easy.

To create a T/PEG 3-layered structure from SWCNT or MWCNT matrix composites the nanophase must be available in P-type and n-type conductors. There are several ways to dope CNTs and in the example discussed here chemical dopants are added to the SWNCT matrix materials and they used to create the n- and p- type matrix composites for the conducting legs of the module structure. Specifically, PEI (polyethylene imide) is added to the nanotube matrix for the doped, n-type leg, and oxygen-doped SWCNT mats have been used for the p-type leg. The specific details of the process are outlined in the references. [18, 19] However, this is for the sake of illustration. These SWCNT thermoelectric contacts could be created in many ways.

The voltage from the p-/n- bilayer with the piezo interposed (a single module layer), is measured across a load resistor placed at the end of the two long sections not in contact with the piezo (where the blue arrows of **Figure 6 (a)** are marked). The DT is applied as shown in the diagram of **Figure 6 (a)**. Comparing this with **Figure 5**, the forcing term is applied in the direction of the polarization of the thin PVDF film.

If we have NO heating at all, the output measured for a sinusoidal forcing term is simply:

$$V_{piezo} = \sigma_{zz} d_z$$

where $s_{zz}$ is the piezoelectric coefficient in the z-direction and $d_z$ is the total dimensional change in that direction. If there is NO forcing, the thermoelectric output for a given DT is:

$$V_{thermo} = \alpha \Delta T$$

where a is the thermoelectric Seebeck coefficient. So, the expected output would simply be a sine wave of amplitude $s_{zz} d_z$ offset by a constant thermovoltage of a$DT$ assuming completely independent energy scavenging.

However, if there is a parametric coupling between the efficiencies these two things don't add together in a straightforward way. Using our simple linear model above we might expect:

$$V_{total} = \sigma'_{zz} d_z + (\alpha' + \alpha_2 \sigma'_{zz} d_z) \Delta T$$

Recall however, that $s_{zz}$ and g may be slightly different from their unperturbed values. In this case we name them with primes. The $a_2$ is the mixing constant introduced above.

In the measurement the forcing term is harmonic so:

$$d_z = A \sin(\omega t)$$

$$V_{total} = A[\sigma'_{zz} + \alpha_2 \sigma'_{zz} \Delta T] \sin(\omega t) + \alpha' \Delta T$$

This would indicate that the amplitude of the oscillatory part of the signal should be temperature gradient dependent. Moreover, the part of the signal specifically associated with the coupling between the thermoelectric and piezoelectric processes (the $a_2$ term). It should be carefully pointed out that this expression does not include the pyroelectric component:

$$V_{pyro} = \frac{A}{R} p_c \frac{dT}{dt}$$

where A is the area of the sample, $p_c$ is the pyroelectric coefficient of PVDF, R is the resistance, and dT/dt is the temperature ramping rate. For PVDF-TrFE, $p_c \sim 40 \ \mu C/m^2 \cdot K$

The effects of the change in overall capacitance when the stack is heated must also be considered. This is because as the system is heated the modulus of the PVDF changes and the change in z-dimension $d_z$ for a given forcing term may change. As a simple linear approximation:

$$C = \frac{\varepsilon A}{d}$$

So:

$$V_{capacitor} = \frac{Q}{\varepsilon A}(d + (d_z + b\Delta T))$$

b in this expression is the rate of change of the strain with DT and $d + d_z$ is the total thickness of the capacitor and this thickness changes by DT as it is heated. From literature we can get the change in the change in the modulus for the PVDF's for the DTs of interest: < 10°C. [20] for both the change in mechanical modulus and the change in $s_{zz}$ with temperature is less than 10%. Moreover, in both cases as temperature increases, these effects give a smaller overall voltage output from the piezo generally.

This means to carry out this experiment and identify the components without confounding voltages we must (1) keep the DT below 10 °C, (2) change DT slowly and measure only after equilibrium has been achieved and (3) compare only across the same forcing frequencies.

Shown in **Figure 7** the p-type / n-type nanotube T/PEG described above, with a $s_{zz}$ -poled PVDF layer is forced. The output voltages are fed directly into a Lacroy capture-scope for two different temperature gradients.

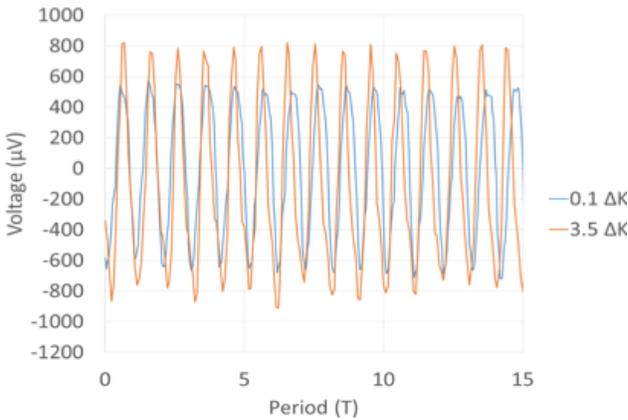

**Figure 7:** The *full output voltage* from a T/PEG is demonstrated here. Using the structure of **Figure 3**, A SWCNT-PVDF T/PEG triple layer was fabricated using O$_2$-doped SWCNTs for the p-type material and PEI doped SWCNTs for the n-type material. In this test, the $\Delta T$ was increased from 0.1 K to 3.5 K, while the oscillatory stress on the structure remained constant. Notice the orange plot is slightly offset in voltage because of the additional thermoelectric voltage. The pyroelectric envelope has fallen to 0 at this time-scale. The increased amplitude is a result of an additional *effect* in the system, and it represents *additional* power generation.

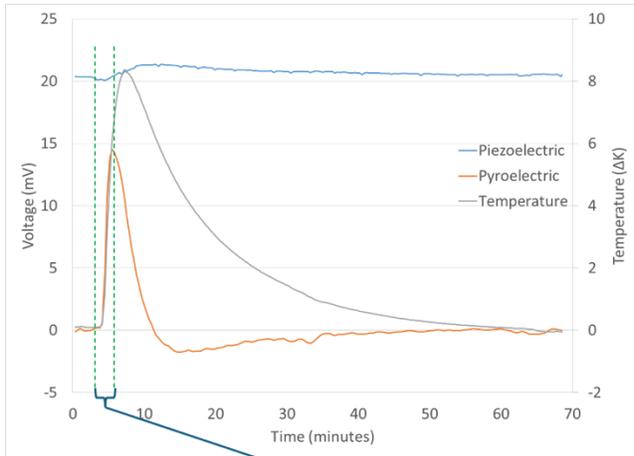

The experiments of **Figure 7** are carried out at 25 ºC and the temperature gradient from that temperature is small. The PVDF film is ~ 100 mm thick. As noted above, the peak-to-peak voltage output is DT dependent and given the conditions discussed above, we expect this to represent only an increase due to the coupling of the thermoelectric and piezoelectric effects. Using the same general structural dimensions, an Al/PVDF/Al stack is used as a control and there is no measurable change with DT in these temperature ranges.

This change in peak-to-peak voltage, together with the pyroelectric response can be seen in **Figure 8** with data taken just as the thermal gradient is established. The top is a 70 min. scan, while below is a 3.5 min. scan. In the top scan the heat is removed after 10 min. and the gradient begins to decrease. The response is seen in the piezo/thermos response (blue) and the pyroelectric response (orange).

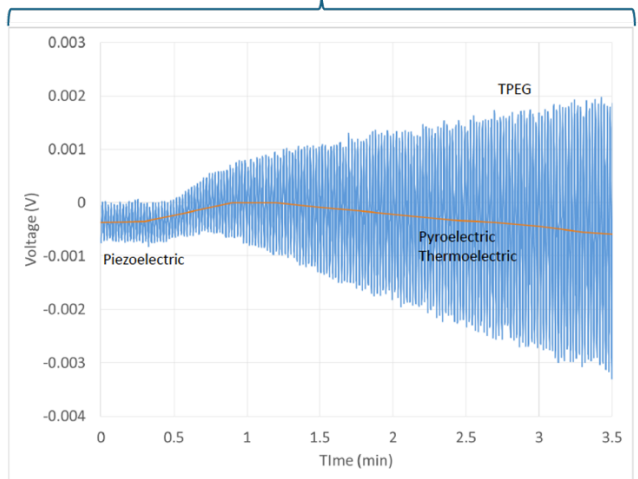

**Figure 8:** The raw data output of the T/PEG as the thermal gradient is established. This allows the pyroelectric component to be separated and avoided. DT = 10 ºC, forcing is 10 Hz, ambient is 25 ºC.

The piezotronic field-doping created by the strained piezo penetrates through the thermoelectric legs subsequently moving the chemical potential (Fermi level colloquially) up or down in the band structure of the nanophase. So, a central mechanism to consider in understanding the observed increase in power can be traced to an increase in Seebeck coefficient. And, the sharper $g(E)$, the greater the value of $d(\ln g(E))/dE$.

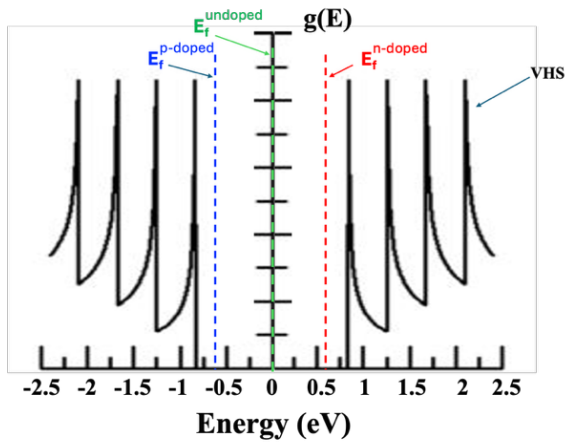

One-dimensional systems have sharp features known as van Hove singularities (VHS) and these repeat across the 1D material's band structure. Moving the Fermi level towards one of these states can be greatly advantageous for the Seebeck coefficient. This is idealized in **Figure 9** where the VHS of a SWCNT is plotted, and the effects of doping are overlaid. [21]

**Figure 9:** an idealized sketch of a 1D band structure with VHS and the position of the Fermi levels as doping conditions change. As doping levels increase in n- or p- directions, the closer the Fermi levels get to the VHS. The density of states at the Fermi level is a factor in the Seebeck coefficient.

Of course, while the Seebeck coefficient is an important characteristic for thermoelectrics, how much power a thermoelectric *can* produce and how much it *does* produce is what must be known. To examine the performance of a heat engine, generally it is necessary to cycle through the thermodynamic variables to see how much work can be done between hot and cold heat baths: as in a Carnot fashion. [22] With the T/PEG, two such heat engines are working together and so it is less obvious how to present such a "work cycle." The clearest analogy is to plot the output power of the T/PEG response across one temperature cycle, as seen in **Figure 10**. Here data of the form as seen in **Figure 8** is collected as the (DT) across the layer's plane is cycled. Ambient temperature is 25 °C. What is displayed is the voltage drop across a load resistor. The load is impedance matched to the T/PEG for current flowing in the direction of the internal diode (see **Figure 6**). Thus, with this load, the power output is maximized for the system and a plot proportional to the total work being done by each of the mechanisms results.

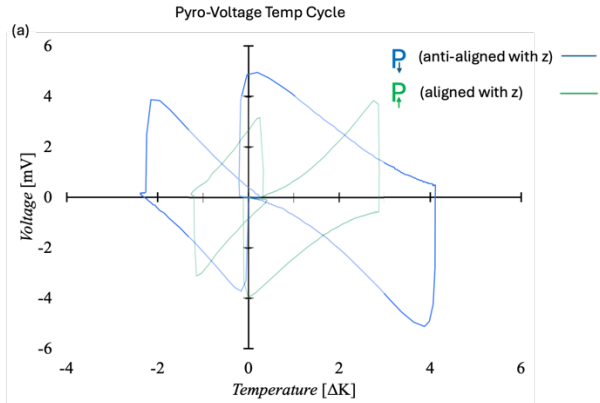

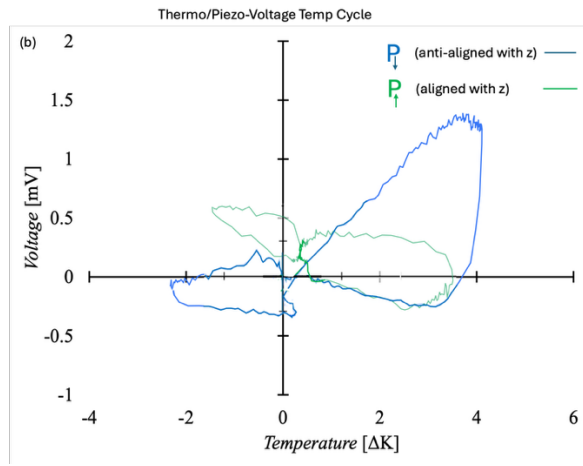

**Figure 10: (a)** This is a temperature cycle plot for the pyroelectric output of the three-layer stack described in **Figure 6(a).** The DT swings from -2K to 4 K around 25 °C. The total area in this case depends on dT/dt which is held constant for both plots. Such plots represent the total work being done in the cycle. **(b)** The temperature cycle plot of the combined thermoelectric-piezoelectric output (see **Figure 8**) is the oscillatory component only. But since the piezo response doesn't change with DT what is captured is the synergistic "additional" component. To examine how this piezotronic field orientation effects the pyro- and entangled- components, the polarization of the inner piezo layer is flipped. The P⁻ indicates that the polarization of the piezo layer is anti-aligned with z which should allow for the largest enhancement as seen in **Figure 4**.

**Figure 10(a)** shows the pyroelectric voltage output for a PVDF:SWCNT system as in **Figure 6**. It is taken at 25 °C. The temperature gradient DT is cycled at a constant rate from -2 K to 4 K and back. The pyroelectric response is roughly symmetric around zero. **Figure 10(b)** shows the thermo/piezo-electric output. It is not symmetric. The asymmetry from positive to negative DT comes from the fact that the system under testing here, has only two thermoelectrically active legs: a p-doped leg and n-doped leg to create a single current path. The junction between the legs forms a diode as in **Figure 4**. The experiment is constructed such that the output voltage is measured across a load resistor that is chosen to maximize the total current flow from the thermoelectric without mechanical forcing. This means it is impedance matched to the internal resistance of the thermoelectric. However, when DT switches direction in this cycle, the current is reversed and flows against the internal diode. The internal potentials are more than enough to exceed the breakdown voltage of this Schottky diode and so current does flow. But the external load resistor is now too small and no longer matches the internal resistance the current encounters. Thus, the voltage output is reduced.

There are two curves shown in each plot, a blue one and a green one. The blue curve represents the case where the direction of the PVDF poling along z (see **Figure 4**) is such that the n-leg gets further "n-doping" from the field, and the p-leg gets "p-doped" from the field. For a positive DT, this increases the thermo-voltage generated in the system for that DT, just as it would if such "doping" had been added chemically. And, since the doping has increased the thermoelectric voltage output under forcing, including the time dependent increase due to synergistic effects, is also increased. Thus, the enclosed area in the blue plot is large. For the green curve, the polarization field points in the opposite direction along z which reduces the overall effect of the chemical dopants used on the n-type/p-type legs. This reduces the thermoelectric voltage by trapping charge, and this is reflected in the area enclosed by the green curve. Notice also that the areas enclosed by the green curves for +/- DT are now nearly the same. This "doping field effect" has reduced the effects of the internal Schottky diode.

Moreover, the cross-over or zero voltage point between the two areas representing +/- DT occurs at zero DT for the blue plots where the piezo poling field adds to the doping effects of the legs, as expected. However, this voltage is surprisingly offset when this field runs the opposite way (the green curves). That means a small thermo-voltage must be generated to overcome some internal voltage generated by the stack when the poling of the piezo placed a field opposite to that of the dopant in the legs. In other words, we have created an interaction between the piezo-field and chemical dopant that traps that dopant's charge in each of the legs. This leads to an internal galvanic effect – like a very small battery. Indeed, this is seen in both the pyroelectric and thermo/piezo-electric green plots of **Figure 10 (a)** and **(b)**.

### 4.2  Chalcogenide Matrix Nanocomposites

T/PEGs, have also used more traditional thermoelectric materials in nanophase forms and these can prove to be significantly more powerful. Thermoelectric properties of small band gap semiconductors are most interesting when the band gap allows for excitation of carriers at the hot end of the DT with extraction of those carriers at the cold end. This reduces back drift of carriers in the system and so increases the ZT. To make full use of such a mechanism, the band structure should be engineered to have a large density of filled states near the Fermi level, a band curvature that allow for fast electrons (small effective masses), small phonon density of states, and so on. If a nanophase matrix composite is used, it is also important to consider hopping dynamics of the charge and how the band structure might play a role in this.

An excellent example of a materials class in which many of these properties are found are the metal chalcogenides. In nanowire and nanoplate formats, chalcogenides are well known to have tunable densities of states and band structures due to the ease of doping and formation of multiple compounds with the same components. They are also easy to synthesize in 1D, 2D, and nanoparticles and are readily dope-able. [23] Binary chalcogenide nanowires can be exceptional thermoelectrics. [24] $Bi_2Te_3$, PbTe, CdTe, $Cu_{2-x}Te$, CdSe, and $Ag_2Se$ nanowires [25] have all been synthesized *via*. low cost, solution-based methods and are widely studied for their thermoelectric properties. Finally, the excellent control over wire morphology usually observed for these systems means material processing to achieve specific matrix composites is particularly advantageous for the T/PEG applications discussed here. However, for the nanophase 1D materials, tendencies to alloy with dopants, and strain induced multi-phase formation, are among the issues that can limit full control of g(E) at the Fermi level, as well as carrier number and mobility.

For this reason, many in the field of thermoelectrics have turned to ternary chalcogenide systems where it is widely held that far more control in such band structure engineering can be achieved. The tunability in ternary chalcogenide alloys comes from an extreme tunability in composition [26], leading to control over the band gap [27], carrier density [28], and photoconductivity. [29] Unfortunately, relatively few classes of metal-Te-Se ternary alloy nanostructures have been synthesized. For the ones that have been demonstrated, control over morphology has been hard to obtain [30], with large diameters, poorly formed phase formation, and aggregation of dopants, generally observed.

Recently, solution-phase synthesis methods for composition-tunable, ultrathin, 1D, ternary metal/Te/Se nanowires have been introduced. This approach produces reliably tune-able properties in the Chalcogenide nanowire systems by using binary chalcogenide nanowire synthesis to create nano-templates. Subsequent post-processing creates the ternary system by dissolution and diffusion under high temperature into the template. This process can be applied to a very broad range of atomic constituents. This rational materials design approach has further been extensively confirmed using calculational *ab initio* methods wherein electron-phonon scattering is combined with a full band structure calculation using density functional theory. Calculations were carried out in the projector augmented wave framework in the Vienna *ab initio* simulation package (VASP). The generalized gradient approximation (GGA) functionals parametrized by Perdew-Burke-Ernzerhof (PBE) was employed. A *self-consistency convergence* criterion of $1\times10^{-6}$ eV and force component tolerance of 0.02 eV/Å was used for these calculations (PBE). These works reveal a striking control over bandgap and fermi level positioning provided by templated synthesis of the nanowires in: Cd-Te-Se, Bi-Te-Se, Ag-Te-Se, Cu-Te-Se, and Pb-Te-Se systems. [31] Among these systems $Bi_2Te_{2.7}Se_{0.3}$ provides an example where ZT ~ 0.75, PF ~ 1.1 X $10^{-3}$ W/cm·$K^2$, s ~ 46500 S/cm all at 325K. [32] These values are for compressed mats of chalcogenide nanowires and are significantly better than what is seen in similar SWCNT morphologies as referenced above.

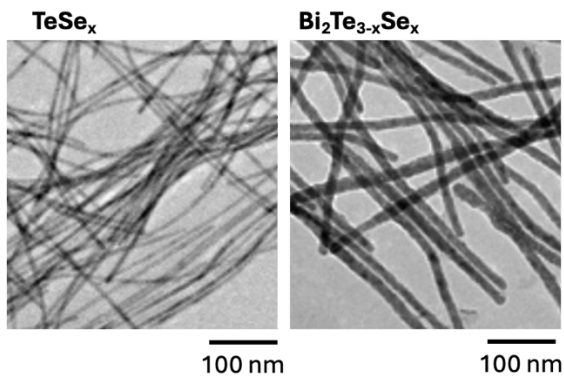

**Figure 11:** a conventional Z-contrast TEM image of the Te-Se, and Bi-Te-Se systems taken at 200 keV on a JEOL 2010F. This demonstrates the overall morphology of the nanowires produced both in binary and ternary materials systems. Growth details are given in the references of [32].

The added dimensionality of plate morphologies can result in significant differences to chalcogenide thermoelectric performance. This has been reported for several of the two-dimensional *nanoplate* analogues of chalcogenide nanowires discussed above. And generally, a very wide variety of the binary and ternary compounds of metal chalcogenides can be made into nanoplates of only a few unit cells thick. [33] In recent years, significant advances have been made in solution-based synthesis methods for these 2D materials and in novel doping schemes which involve placing metallic dopants around the edge of the plates in a modulation doping or remote doping analogue. [34] As with the nanowires of above, ultra-thin, metal-nanoparticle-doped, chalcogenide plates have been extensively studied with *ab initio* methods. And, just as with nanowire systems, an astonishingly good correspondence can be found between density functional approaches and the measured band characteristics in the chalcogenide plates. [see 33]

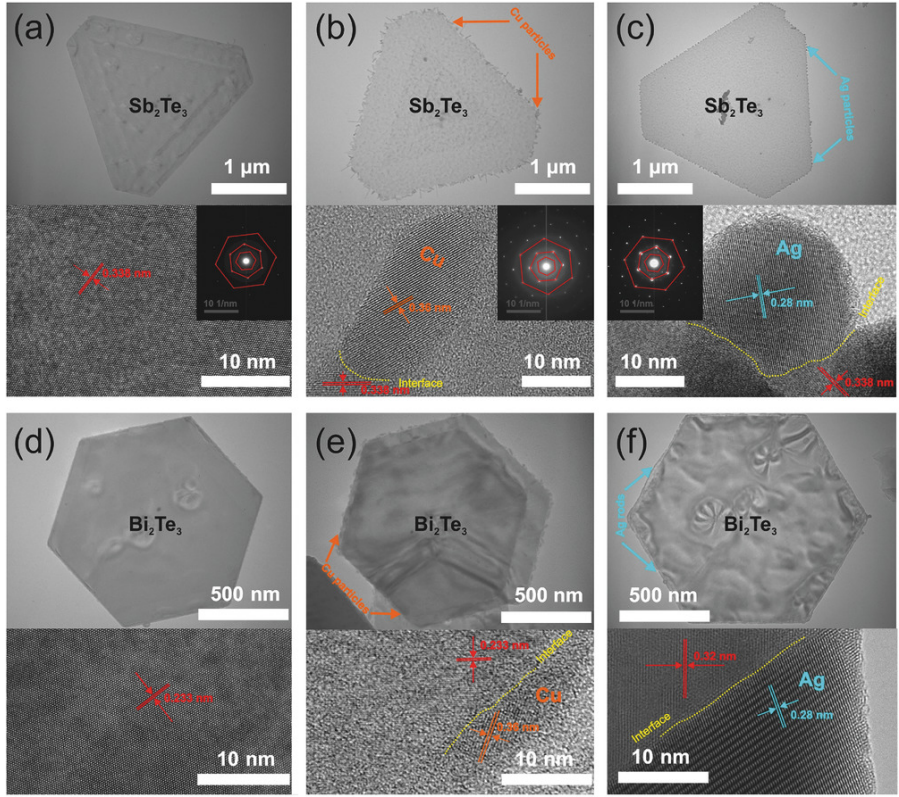

**Figure 12:** TEM and HRTEM images of the lateral 2D heterostructures based on $V_2$–$VI_3$ nanoplates with different self-assembled metal nanoparticles. Chalcogenides a wide toolbox of composition and dopants. Imaged with JEOL 2010F STEM at 200 keV.

What has been exciting is that the doping effect on such plates depends starkly on the dopant's position with edge doping (see **Figure 12**) adding to the thermoelectric characteristics and doping at planar or surface sites significantly reducing thermoelectric figures of merit. This has been conjectured to arise from the local alloying of planar or surface species to modify the bands locally introducing strong scattering whereas edge sites do not do this and instead introduce free electrons into the band structure enhancing conductivity. [33] Indeed, doped 2D plates can show significant increases in power output as compared to nanowire or undoped 2D plate counterparts, both in terms of their thermoelectric power and their T/PEG power.

Importantly, spin-orbit coupling in many of these 2D chalcogenide systems are well known to produce symmetry protected states or *topological* states. It is conjectured that such states are always associated with the boundaries of the system and express a dramatic robustness to electron scattering. It has already been suggested that such states could an important role in modifications to Seebeck coefficients for 2D doped metal chalcogenide thermoelectrics. [35] Hints of this can be seen for magnetic dopants as in the case of **Figure 12** where $Sb_2Te_3$ is doped with both Fe and Ag and the two compared for the thermoelectric output during T/PEG driving as in the configuration of **Figure 4**.

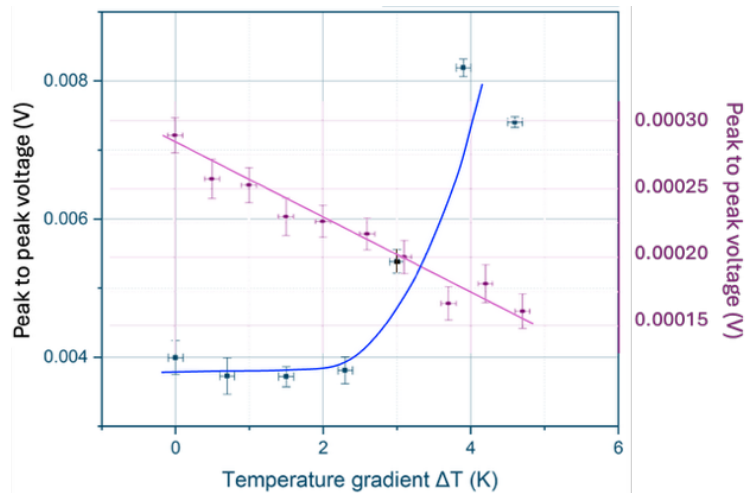

**Figure 13:** a comparison between Ag and Fe doped $Sb_2Te_3$. Notice the curious behavior of the Fe-doped materials (purple). The peak top peak voltage from the thermoelectric output is an order of magnitude smaller than for the Ag doped materials (blue). Moreover, the trend of increasing thermoelectric voltage with DT as seen in most other materials and in the case of Ag here, is not seen for Fe. A conjectural explanation for this is that the Fe is magnetic, and the temperature increase changes the scattering of the current carried in the topological edge states of the $Sb_2Te_3$.

Of course, this is not a full review of nano-chalcogenide synthesis, but it should be clear that recent advances in synthesis have made morphological and electronic control over this thermoelectrically important class of materials far more obtainable. Specifically rational design of doped binary and ternary compounds into a wide variety of morphological variants has been a striking achievement of solvo-chemical synthesis over the past 10 years allowing for explorations of the interplay between band structure and the complex confinement geometries of shaped 2D manifolds.

Returning again to the testing configuration of **Figure 4**, wherein a piezo-layer is interposed between a two thermoelectric leg "module," the comparative performance of a range of compositions can be made. For such a comparison the nanophase loading with respect to percolation in each matrix composite is held constant. In this case the average nanowire lengths and aspect ratios are roughly equal. Forcing terms and DT's are all equal, as are the physical dimensions and nanophase dispersion characteristics as determined by light scattering. **Figure 14** shows the power factors (blue bars) and the static ZT (green bars). The ZT results when calculated during forcing, that is the "kinetic" ZT or $ZT_k$ are shown in the cross hatched bars on top of the static values. So, for example, the $ZT_k$ for Cu doped $Bi_2Te_3$ plates is ~ 3! For PEI/O doped SWCNTs $ZT_k$ approaches 0.5 whereas the static ZT's for these materials are 1 and 0.25 respectively.

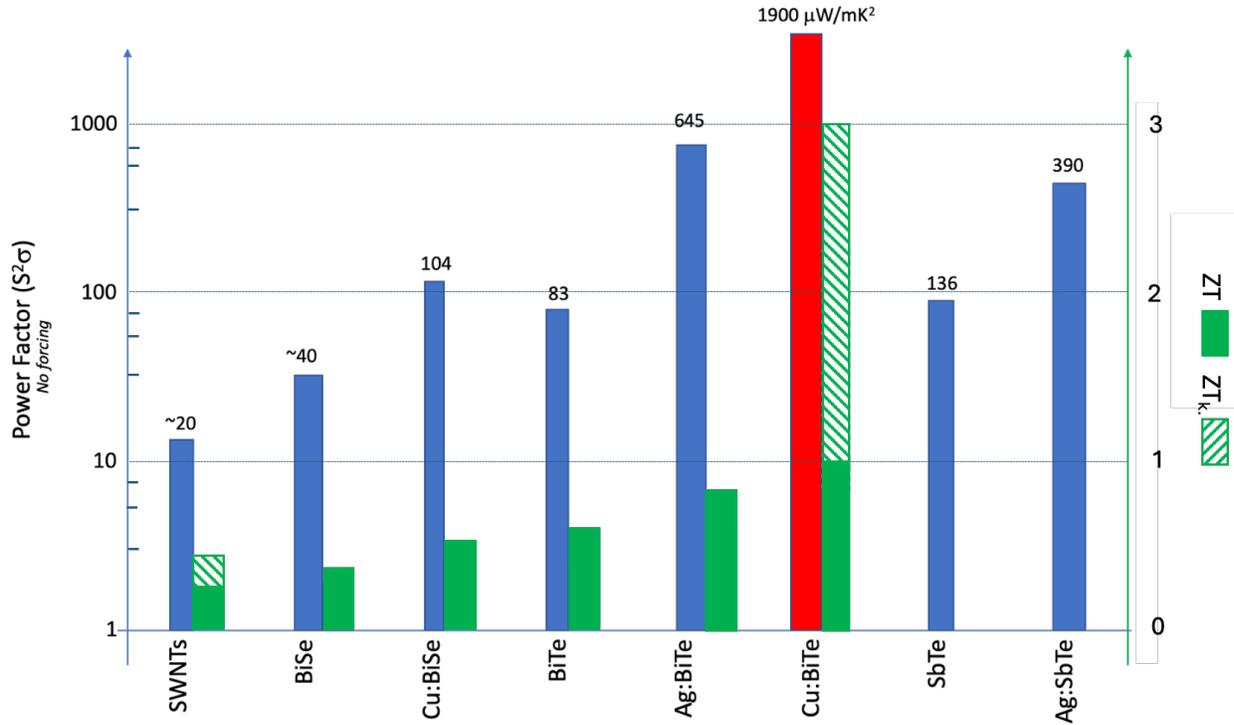

**Figure 14**: A comparison of the figures of merit as discussed in the text, for a variety of compounds found in literature. The chalcogenides listed here are in plate format, the SWCNT devices are used as a comparison. The solid green lines are for static ZT whereas the green cross hatched bars are the effective ZT under loading ($ZT_k$). The numbers for the Power factors are given in blue with the red highlighting the largest number. The PF is taken for the unforced systems, so as can be seen the Cu:BiTe system is quite large for any flexible thermoelectric.

So, what is $ZT_k$ and what relevance does it have to T/PEG performance? $ZT = \alpha^2\sigma T/\kappa$ where, under forcing, T and k generally remain the same as in the static case. So, it is the power factor that is changing. Both the Seebeck coefficient and the conductivity will change under forcing. But keep in mind that the load resistance across which this is being measured is also a constant. The Seebeck and conductivity change implies a change in the internal impedance of the stack. Thus, the load conditions are no longer the same as they were for the static case, and this suggests a change in peak power output just on the basis of impedance mismatch. This means direct interpretation of comparisons between $ZT_k$'s can be tricky. As seen in **Figure 13**, the static ZT's of Cu:BiTe > Ag:BiTe by only about 25%. However, while the static PF of Ag:BiTe is ~ 645 the static PF of the Cu:BiTe is nearly 4X larger (shown in red). This increase is reflected by a 3X increase in $ZT_k$ from ZT of the Cu:BiTe system.

The "additional" power a thermal/kinetic activated T/PEG assembly will generate varies from material to material used in the TE legs. There appears to be a direct correlation between the power factor (PF) of the TEG component and this additional power as seen in **Figure 15**. This is likely due to the specifics of the band structure to which the piezo-field has coupled.

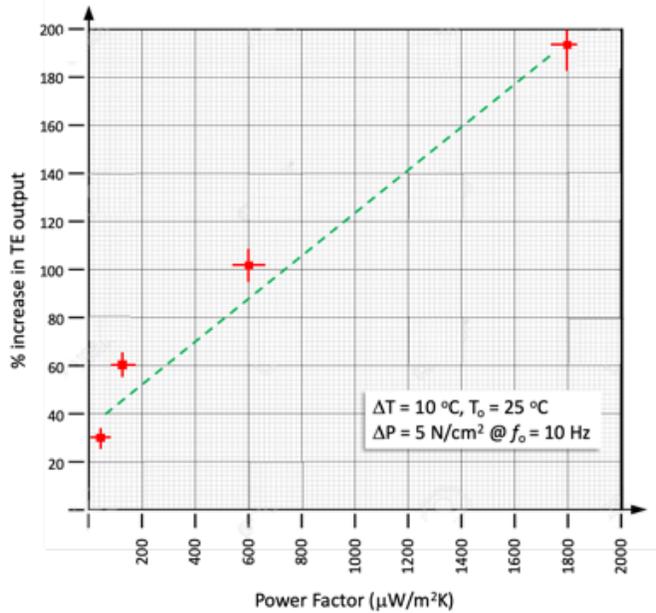

**Figure 15:** The percentage of power increase in a T/PEG system depends on how good of a thermoelectric is used. Here we compare the amount of additional thermally-related power generated (for the mechanical driving conditions given) as we increase the power factor of the device. Specifically, four devices were built using different materials for the thermoelectric legs: SWNTs, and doped dichalcogenides. Different doping levels in the dichalcogenides yielded different power factors in the thermoelectric as one would expect. Each material stack was then tested under a driving compression of DP = 5N/cm$^2$ @ 10 Hz (sine wave). The change in the output power was determined as a percentage of the total output power without driving: Dx/x$_{DP=0}$.

## 5.0. Scavenging Power to Cool: Combining Peltier with Electrocalorics

For thermoelectric systems, Onsager's reciprocal relations suggest that the Seebeck effect and the Peltier effect are directly related to each other. The Seebeck effect being the observed V$_{th}$ across the materials with the application of a DT and the Peltier effect being the observation of a DT with the application of a Voltage across the material. Similarly, these same reciprocal relations connect piezoelectricity and electroactuation in the same way – one the creation of a Voltage due to applied stress, and the other the dimensional expansion of contraction of a volume of material with the application of a voltage. [36]

For single effect systems Onsager's analysis is quite simple to apply. And when two different effects exist in the same system, but they are not "entangled" thermodynamically, again the application of reciprocal effects is completely separable. However, when a system expresses two different thermodynamic effects and those effects couple together, as in the T/PEG architecture, this coupling can lead to some ambiguity of our expectations as to reciprocal effects. The T/PEG's synergistic effects present a particularly difficult problem in this regard. To get to the reciprocal effect dynamical Electrocalorics and Peltier effects become mixed. [37]

To see why this is, we use the three-layer device of **Figure 15** to illustrate. In **Figure 15** a 2-SbTe-leg T/PEG using a PVDF interlayer (three layers total) is configured in a circuit geometry wherein the load, R$_{load}$ is small. The drive is between 50 and 500 Hz with a 3 cm$^2$ forcing area and a 5 N force giving a pressure of 16666.67 Pa. The measured temperature shift from ambient (25 °C) is shown as a function of driving frequency. Importantly this is measured below the body of the layered sample. The load resistance heats up and this represents cooling over the whole 3 cm$^2$ sample area. Thus, we have created a thermal pump.

The origin of this thermal pump is twofold: clearly there is a small kinetic electrocaloric effect in operation. However, secondly, there is a Peltier effect due to the applied current in the thermodynamic legs when driving is present. The DT seen in the graph at the left is the addition of those two effects. It should be emphasized that this Peltier-Caloric Effect (PCE) here is not optimized for maximum cooling, but its observation does suggest that the thermoelectric and piezoelectric coupling may have deeper implications.

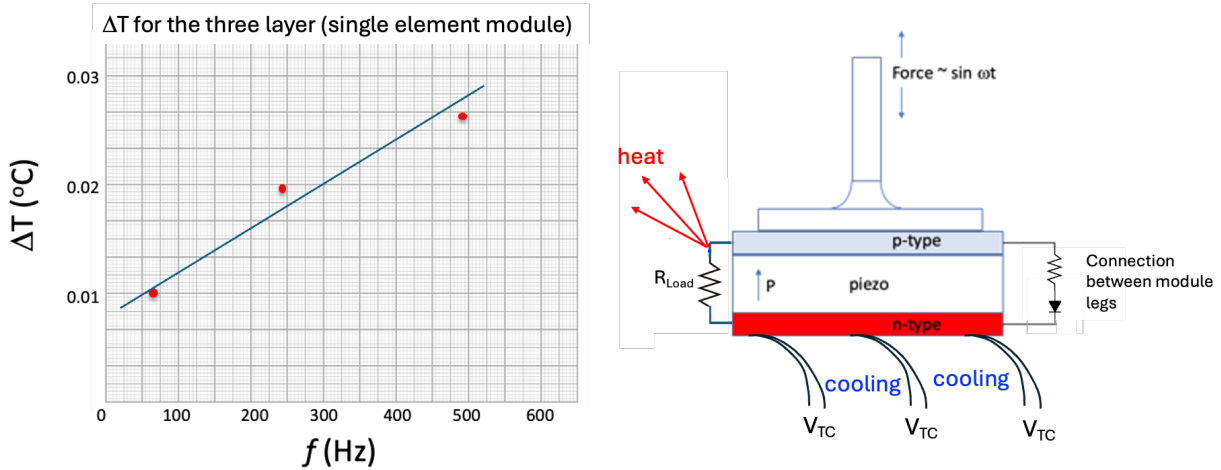

**Figure 15:** Measured with a K-type thermocouple with a 32 bit ADC and a 10V range. Measurements are +/- 0.000057 °C. The TC is embedded in the support substrate below T/PEG unit in test with it making direct contact. This represents a direct measurement of the kinetic electrocaloric effect for one module (two legs) of a T/PEG. The frequency of forcing is plotted as a function of the DT that the forcing creates. Importantly the forcing is a relatively small.

Even though the PCE demonstrated here is small, it is NOT local. This temperature change is measured across the whole sample surface. For the 3 layers of this device across 3 cm$^2$ and a forcing motion of only microns, the total cooling power Can be estimated at 500 Hz as roughly 0.35 W +/- 0.1W, with the uncertainty coming in the timing. For such a thin layered device that has not been optimized for maximum DT and is not fully isolated from the environment, this is an astonishing level of cooling. Indeed, it is in the same orders of magnitude as much more sophisticated, and optimized ceramic devices. [38]

## 6.0 Conclusions

The T/PEG is a combined thermoelectric/piezoelectric power generator which scavenges heat and kinetic energy and converts it into an electrical power output. It is basically constructed by using thermoelectrically active, matrix nanocomposite electrodes - in which the nanophase is a reduced dimensional small bandgap semiconductor - placed onto a thin, poled layer of piezoelectric forming a multilayered structure of only 100 microns or so. By allowing for sufficient piezo-field penetration into the volume of the thermoelectric matrix, the chemical potential of the nanophase can be adjusted toward high electron density of state positions of the nanophase band structure (when such features exist), thereby increasing the Seebeck coefficient, and the conductivity of the nanophase significantly. This "field-doping" during forcing of the piezoelectric allows the thermoelectric power generation to be significantly enhanced at the cost of the work done by the piezo.

A coupled heat engine description of the T/PEG produces a reasonable approach to explaining the observed power generated by these devices. Moreover, it suggests that the significant power generation increase as compared to the two modalities independently may come from several sources including field doping, load impedance matching, and piezo-field alignment within the matrix. With the currently measured power profiles, T/PEGs offer an interesting new route toward ambient power scavenging in technologies ranging from transport to wearables. Even more exciting is the promise of new phenomena as predicted through Onsager reciprocity such as the kinetic PCE in large scale coverings and coatings.

**Acknowledgements**

The authors thankfully acknowledge D. Montgomery, C. Hewitt, H.H. Huang, and S. Roth for their many contributions to understanding these effects. Funding for this work was provided by AFOSR grant no. FA9550-16-1-0328